\documentclass[thmsa,11pt]{article}
\usepackage{amsfonts}

\usepackage{amsmath}
\usepackage{graphicx}

\input{tcilatex}

\begin{document}

\setcounter{page}{0} \topmargin0pt \oddsidemargin5mm \renewcommand{%
\thefootnote}{\fnsymbol{footnote}} \newpage \setcounter{page}{0} 
\begin{titlepage}
\begin{flushright}
\end{flushright}
\vspace{0.5cm}
\begin{center}
{\Large {\bf The twisted XXZ chain at roots of unity revisited} }

\vspace{0.8cm}
{ \large Christian Korff}

\vspace{0.5cm}
{\em School of Mathematics, University of Edinburgh\\
Mayf{i}eld Road, Edinburgh EH9 3JZ, UK}
\end{center}
\vspace{0.2cm}
 
\renewcommand{\thefootnote}{\arabic{footnote}}
\setcounter{footnote}{0}

\begin{abstract}
The symmetries of the twisted XXZ chain alias the six-vertex model at roots of unity are 
investigated. It is shown that when the twist parameter is chosen to depend on the total spin an 
inf{i}nite-dimensional non-abelian symmetry algebra can be explicitly constructed for all 
spin sectors. This symmetry algebra is identif{i}ed to be the upper or lower Borel 
subalgebra of the $sl_2$ loop algebra. The proof uses only the intertwining property of the 
six-vertex monodromy matrix and the familiar relations of the six-vertex Yang-Baxter algebra. 

\medskip
\par\noindent
\end{abstract}
\vfill{ \hspace*{-9mm}
\begin{tabular}{l}
\rule{6 cm}{0.05 mm}\\
c.korff@ed.ac.uk
\end{tabular}}
\end{titlepage}
\newpage

\section{Introduction}

In recent years there has been renewed interest in the degeneracies
exhibited by the integrable six-vertex model and the associated XXZ quantum
spin-chain, 
\begin{equation}
H=\sum_{m=1}^{M}\sigma _{m}^{+}\sigma _{m+1}^{-}+\sigma _{m}^{-}\sigma
_{m+1}^{+}+\frac{q+q^{-1}}{4}\,\sigma _{m}^{z}\sigma _{m+1}^{z},\quad \sigma
_{M+1}^{\pm }\equiv \sigma _{1}^{\pm },\;\sigma _{M+1}^{z}\equiv \sigma
_{1}^{z}\,,  \label{H}
\end{equation}
when the anisotropy parameter is evaluated at roots of unity $q^{N}=1$. In 
\cite{DFM} Deguchi, Fabricius and McCoy showed for the commensurate sectors $%
2S^{z}=0\func{mod}N$ (with $S^{z}$ being the total spin) that the model with
periodic boundary conditions exhibits an $\widetilde{sl}_{2}=sl_{2}\otimes 
\mathbb{C}[t,t^{-1}]$ loop algebra symmetry. Outside these commensurate
sectors the algebraic structure of the symmetry algebra has so far not been
established except for the case of the XX model, i.e. vanishing anisotropy
parameter, and a numerical construction for $N=3,$ see \cite{DFM} for
details.

In two recent works \cite{D03,T03} the twisted XXZ chain at roots of unity
has been investigated, 
\begin{equation}
H^{\lambda }=\sum_{m=1}^{M}\sigma _{m}^{+}\sigma _{m+1}^{-}+\sigma
_{m}^{-}\sigma _{m+1}^{+}+\frac{q+q^{-1}}{4}\,\sigma _{m}^{z}\sigma
_{m+1}^{z},\quad \sigma _{M+1}^{\pm }\equiv \lambda ^{\pm 1}\sigma _{1}^{\pm
},\;\sigma _{M+1}^{z}\equiv \sigma _{1}^{z}\;.  \label{Ham}
\end{equation}
In \cite{D03} various operators have been constructed which (anti)commute
with the twisted XXZ Hamiltonian and the associated transfer matrix. Except
for the cases $\lambda =-1$ the algebraic structure underlying these
operators has not been identif{i}ed. Similar as for the periodic case the $%
\widetilde{sl}_{2}$ symmetry algebra for $\lambda =-1$ has been restricted
to certain commensurate spin-sectors. The discussion in \cite{D03} has also
been extended to include the case of the inhomogeneous chain.

In the second work \cite{T03} the construction of operators creating
complete strings for the periodic homogeneous chain carried out in \cite
{FM01b} has been generalised to cover also the twisted and inhomogeneous
case. The construction of the symmetry algebra underlying the degeneracies
in the spectra of the Hamiltonian and the transfer matrix has not been
investigated.

In this letter it is shown that when the twist parameter is chosen to depend
on the total spin, i.e. $\lambda =q^{\pm 2S^{z}}$, the quantum spin chain
Hamiltonian and the associated twisted six-vertex transfer matrix exhibit inf%
{i}nite-dimensional non-abelian symmetries and their algebraic structure is
identif{i}ed with the lower respectively upper Borel subalgebra $U(b_{\mp
})\subset U(\widetilde{sl}_{2})$ in \emph{all} spin sectors. In the sectors $%
2S^{z}=0\func{mod}N$ one obviously recovers the periodic chain and the
symmetry is enhanced to the full loop algebra $U(\widetilde{sl}_{2})$
reproducing the aforementioned result of \cite{DFM}. However, also for the
periodic case we give a novel proof of the symmetry which only uses the
framework of the algebraic Bethe ansatz \cite{QISM} and quantum group theory 
\cite{Jimbo,Drin0}. In particular, it avoids having f{i}rst to prove
translation invariance, cf. \cite{DFM,KM01,KR02}. The extension to the
inhomogeneous case is also discussed.

\section{The twisted six-vertex model}

Starting point of our discussion is the six-vertex $R$-matrix which is given
by 
\begin{equation}
R(z,q)=\tfrac{a+b}{2}1\otimes 1+\tfrac{a-b}{2}\,\sigma ^{z}\otimes \sigma
^{z}+c\,\sigma ^{+}\otimes \sigma ^{-}+c^{\prime }\sigma ^{-}\otimes \sigma
^{+}  \label{R}
\end{equation}
where we choose the following parametrization of the Boltzmann weights 
\begin{equation}
a=1,\quad b=\frac{1-z}{1-zq^{2}}\,q,\quad c=\frac{1-q^{2}}{1-zq^{2}},\quad
c^{\prime }=c\,z\;.  \label{abc}
\end{equation}
Here $z$ denotes the (multiplicative) spectral parameter and $q$ is the
deformation parameter appearing in the spin-chain Hamiltonians (\ref{H}) and
(\ref{Ham}). Central to our discussion will be the properties of the
(inhomogeneous) six-vertex monodromy matrix which one usually decomposes
over the two-dimensional auxiliary space, 
\begin{equation}
R_{0M}(z/\zeta _{M})\cdots R_{01}(z/\zeta _{1})=\sigma ^{+}\sigma
^{-}\otimes A+\sigma ^{+}\otimes B+\sigma ^{-}\otimes C+\sigma ^{-}\sigma
^{+}\otimes D\;.  \label{mom}
\end{equation}
The explicit dependence on the spectral parameter and the inhomogeneity
parameters $\zeta =(\zeta _{1},...,\zeta _{M})$ will be often suppressed in
the notation in order to unburden the formulas. The twisted six-vertex
transfer matrix is now def{i}ned as the trace 
\begin{equation}
T^{\lambda }(z)=\limfunc{Tr}_{0}\lambda ^{\frac{\sigma ^{z}\otimes \mathbf{1}%
}{2}}\,R_{0M}(z/\zeta _{M})\cdots R_{01}(z/\zeta _{1})=\lambda ^{\frac{1}{2}%
}\,A(z)+\lambda ^{-\frac{1}{2}}D(z)\;.  \label{T}
\end{equation}
For the homogeneous chain $\zeta _{1}=\cdots =\zeta _{M}=1$ we obtain up to
an additive constant the spin-chain Hamiltonian (\ref{Ham}) as the following
logarithmic derivative 
\begin{equation}
H^{\lambda }=(q-q^{-1})\;T^{\lambda }(z)^{-1}\left. z\frac{d}{dz}T^{\lambda
}(z)\right| _{z=1}+M\,\frac{q+q^{-1}}{2}\;.
\end{equation}
Obviously, the twist does not alter the algebraic relations of the
Yang-Baxter algebra def{i}ned in terms of $\{A,B,C,D\}$ in (\ref{mom}). In
order to discuss the symmetries of (\ref{Ham}) and (\ref{T}) when the
deformation parameter $q$ is a root of unity we f{i}rst establish a number
of relations between the Chevalley-Serre basis of the quantum group $U_{q}(%
\widetilde{sl}_{2})$ and the matrix elements of the monodromy matrix (\ref
{mom}) for generic $q$ and $\lambda $.

\section{The Chevalley-Serre basis of $U_{q}(\widetilde{sl}_{2})$}

It is well known that the underlying algebraic structure of the six-vertex
model is the quantum loop algebra $U_{q}(\widetilde{sl}_{2})$. Its algebraic
def{i}nition \cite{Jimbo,Drin0} in terms of the Chevalley-Serre basis is 
\begin{equation}
k_{i}e_{j}k_{i}^{-1}=q^{\mathcal{A}_{ij}}e_{j},\quad
k_{i}f_{j}k_{i}^{-1}=q^{-\mathcal{A}_{ij}}f_{j},\quad
k_{i}k_{j}=k_{j}k_{i},\quad i,j=0,1  \label{AQG}
\end{equation}
where the Cartan matrix reads 
\begin{equation*}
\mathcal{A}=\left( 
\begin{array}{cc}
2 & -2 \\ 
-2 & 2
\end{array}
\right) \;.
\end{equation*}
In addition one has to impose for $i\neq j$ the Chevalley-Serre relations, 
\begin{eqnarray}
e_{i}^{3}e_{j}-[3]_{q}e_{i}^{2}e_{j}e_{i}+[3]_{q}e_{i}e_{j}e_{i}^{2}-e_{j}e_{i}^{3} &=&0
\notag \\
f_{i}^{3}f_{j}-[3]_{q}f_{i}^{2}f_{j}f_{i}+[3]_{q}f_{i}f_{j}f_{i}^{2}-f_{j}f_{i}^{3} &=&0\;.
\label{CS}
\end{eqnarray}
The quantum algebra can be made into a Hopf algebra upon def{i}ning a
coproduct which we choose to be 
\begin{equation}
\Delta (e_{i})=1\otimes e_{i}+k_{i}\otimes e_{i},\;\Delta
(f_{i})=f_{i}\otimes k_{i}^{-1}+1\otimes f_{i},\;\Delta (k_{i})=k_{i}\otimes
k_{i},\;i=0,1\,.  \label{cop}
\end{equation}
The opposite coproduct $\Delta ^{\text{op}}$ is obtained by permuting the
two factors. The six-vertex $R$-matrix intertwines the two coproduct
structures in the case of the spin $1/2$ representation, i.e. 
\begin{equation}
R(z/\zeta )(\pi _{z}\otimes \pi _{\zeta })\Delta (x)=(\pi _{z}\otimes \pi
_{\zeta })\Delta ^{\text{op}}(x)R(z/\zeta )  \label{int}
\end{equation}
with the representation $\pi _{z}:U_{q}(\widetilde{sl}_{2})\rightarrow 
\limfunc{End}\mathbb{C}^{2}$ given in terms of Pauli matrices by 
\begin{eqnarray}
\pi _{z}(e_{0}) &=&z\sigma ^{-},\quad \pi _{z}(f_{0})=z^{-1}\sigma
^{+},\quad \pi _{z}(k_{0})=q^{-\sigma ^{z}}  \notag \\
\pi _{z}(e_{1}) &=&\sigma ^{+},\quad \pi _{z}(f_{1})=\sigma ^{-},\quad \pi
_{z}(k_{1})=q^{\sigma ^{z}}\;.  \label{pi0}
\end{eqnarray}
From the fusion relation $(1\otimes \Delta )R=R_{13}R_{12}$ an analogous
intertwining relation follows for the monodromy matrix (\ref{mom}) with
regard to the quantum group generators on the quantum spin-chain $\pi
_{\zeta _{1}}\otimes \cdots \otimes \pi _{\zeta _{M}}$, 
\begin{eqnarray}
K_{i} &=&q^{\varepsilon _{i}\sigma ^{z}}\otimes \cdots \otimes
q^{\varepsilon _{i}\sigma ^{z}}=q^{\varepsilon _{i}2S^{z}}  \notag \\
E_{i} &=&\sum_{m=1}^{M}\zeta _{m}^{\delta _{i0}}\,q^{\varepsilon _{i}\sigma
^{z}}\otimes \cdots q^{\varepsilon _{i}\sigma ^{z}}\otimes \underset{m^{%
\text{th}}}{\sigma ^{\varepsilon _{i}}}\otimes 1\cdots \otimes 1\;  \notag \\
F_{i} &=&\sum_{m=1}^{M}\zeta _{m}^{-\delta _{i0}}\,1\otimes \cdots 1\otimes 
\underset{m^{\text{th}}}{\sigma ^{-\varepsilon _{i}}}\otimes q^{-\varepsilon
_{i}\sigma ^{z}}\cdots \otimes q^{-\varepsilon _{i}\sigma ^{z}},\quad
\varepsilon _{i}:=(-1)^{i+1}\;.  \label{KEF}
\end{eqnarray}
Here $i=0,1$ as before\footnote{%
Notice that we have chosen to work in the homogeneous gradation in (\ref{pi0}%
) in accordance with the parametrization (\ref{abc}) of the Boltzmann
weights. Equally well, we could have used the principal gradation in which
the six-vertex R-matrix (\ref{R}) is symmetric. Then all Chevalley-Serre
generators in (\ref{pi0}) would acquire a spectral parameter dependence and
the generators (\ref{KEF}) would correspond to those discussed in equation
(50) of \cite{D03}. The choice of the gradation does not alter the algebraic
structure.}. From the intertwining property of the monodromy matrix one then
obtains the commutators 
\begin{equation}
\lbrack
A,K_{1}]=[D,K_{1}]=0,\;K_{1}BK_{1}^{-1}=q^{-2}B,\;K_{1}CK_{1}^{-1}=q^{2}C
\label{KABCD}
\end{equation}
and 
\begin{eqnarray}
\left[ E_{1},A\right] _{q} &=&-K_{1}C,\;\left[ E_{1},B\right]
_{q^{-1}}=A-K_{1}D,\;\left[ E_{1},C\right] _{q}=0,\;\left[ E_{1},D\right]
_{q^{-1}}=C  \notag \\
\left[ A,F_{1}\right] _{q^{-1}} &=&-BK_{1}^{-1},\;\left[ B,F_{1}\right]
_{q^{-1}}=0,\;\left[ C,F_{1}\right] _{q}=A-DK_{1}^{-1},\;\left[ D,F_{1}%
\right] _{q}=B\;.  \label{ABCD}
\end{eqnarray}
Here $[x,y]_{q}=xy-q\,yx$. The commutation relations for the aff{i}ne
generators $\{E_{0},F_{0},K_{0}\}$ are obtained by the simultaneous
replacement 
\begin{equation}
(A,B,C,D)\rightarrow (D,z^{-1}C,zB,A)\quad \text{and\quad }%
(E_{1},F_{1},K_{1})\rightarrow (E_{0},F_{0},K_{0})\;.  \label{auto}
\end{equation}
Note that for the homogeneous case $\zeta _{1}=\cdots =\zeta _{M}=1$ the
above algebra automorphism is implemented by the spin-reversal operator $%
\frak{R}=\sigma ^{x}\otimes \cdots \otimes \sigma ^{x}$, 
\begin{equation}
\frak{R\,}E_{i}\frak{R}=E_{i+1},\quad \frak{R\,}F_{i}\frak{R}=F_{i+1},\quad 
\frak{R\,}K_{i}\frak{R}=K_{i+1},\quad i\in \mathbb{Z}_{2}\;.
\end{equation}

Instead of the spin $1/2$ representation (\ref{pi0}) one might of course
equally well use evaluation representations of higher spin in the definition
of the spin-chain generators (\ref{KEF}), similar as it has been done in 
\cite{T03}. As long as the auxiliary space is not altered the form of the
commutation relations (\ref{KABCD}) and (\ref{ABCD}) is unchanged. From (\ref
{ABCD}) one now deduces by a straightforward computation the following
relations for the twisted six-vertex transfer matrix 
\begin{eqnarray}
E_{1}^{n}T^{\lambda } &=&(q^{n}\lambda ^{\frac{1}{2}}A+q^{-n}\lambda ^{-%
\frac{1}{2}}D)E_{1}^{n}+\lambda ^{-\frac{1}{2}}[n]_{q}(1-\lambda
K_{1})CE_{1}^{n-1},  \notag \\
E_{0}^{n}T^{\lambda } &=&(q^{-n}\lambda ^{\frac{1}{2}}A+q^{n}\lambda ^{-%
\frac{1}{2}}D)E_{0}^{n}+\lambda ^{\frac{1}{2}}\,z[n]_{q}(1-\lambda
^{-1}K_{0})BE_{0}^{n-1}  \label{Tb+}
\end{eqnarray}
and 
\begin{eqnarray}
F_{1}^{n}T^{\lambda } &=&(q^{n}\lambda ^{\frac{1}{2}}\,A+q^{-n}\lambda ^{-%
\frac{1}{2}}D)F_{1}^{n}-\lambda ^{-\frac{1}{2}}[n]_{q}q^{-n}(1-\lambda
K_{1}^{-1})F_{1}^{n-1}B,  \notag \\
F_{0}^{n}T^{\lambda } &=&(q^{-n}\lambda ^{\frac{1}{2}}\,A+q^{n}\lambda ^{-%
\frac{1}{2}}D)F_{0}^{n}-\lambda ^{\frac{1}{2}}\,z^{-1}[n]_{q}q^{-n}(1-%
\lambda ^{-1}K_{0}^{-1})F_{0}^{n-1}C\;.  \label{Tb-}
\end{eqnarray}
We are now in the position to discuss the symmetry algebras of the twisted
six-vertex transfer matrix at roots of unity.

\section{Inf{i}nite non-abelian symmetries at $q^{N}=1$}

Henceforth we set the deformation parameter $q$ to be a primitive root of
unity of order $N\geq 3$. This entails signif{i}cant changes in the
algebraic structure of the quantum loop algebra $U_{q}(\widetilde{sl}_{2})$.
There now exist two versions of the algebra, one of them, which we keep
denoting by $U_{q}(\widetilde{sl}_{2}),$ has an enlarged centre compared to
generic $q$. Its representation theory has been discussed to some extent in 
\cite{BK}. The other version from which we will obtain the symmetry
generators is the restricted quantum algebra $U_{q}^{\text{res}}(\widetilde{%
sl}_{2})$. It can be realised as automorphisms over $U_{q}(\widetilde{sl}%
_{2})$. Details on its representation theory can be found in \cite{CP}. For
the present purposes it will be important that for evaluation
representations of the form (\ref{pi0}) used in the def{i}nition of the
quantum spin-chain one can write down explicit formulas for the generators
of $U_{q}^{\text{res}}(\widetilde{sl}_{2}):$ for some $\tilde{q}$ with $%
\tilde{q}^{N}\neq 1$ and $n\in \mathbb{N}$ we set 
\begin{multline*}
E_{1}^{(n)}=\lim_{\tilde{q}\rightarrow q}E_{1}^{n}(\tilde{q})/[n]_{\tilde{q}%
}!= \\
\tsum_{m_{i}}\;q^{n\sigma ^{z}}\otimes \cdots \otimes \underset{m_{1}^{\text{%
th}}}{\sigma ^{+}}\otimes q^{(n-1)\sigma ^{z}}\cdots \otimes \underset{%
m_{2}^{\text{th}}}{\sigma ^{+}}\otimes q^{(n-2)\sigma ^{z}}\cdots q^{\sigma
^{z}}\otimes \underset{m_{n}^{\text{th}}}{\sigma ^{+}}\otimes 1\cdots
\otimes 1
\end{multline*}
\begin{multline*}
E_{0}^{(n)}=\lim_{\tilde{q}\rightarrow q}E_{0}^{n}(\tilde{q})/[n]_{\tilde{q}%
}!= \\
\tsum_{m_{i}}\zeta _{m_{1}}\cdots \zeta _{m_{n}}q^{-n\sigma ^{z}}\otimes
\cdots \otimes \underset{m_{1}^{\text{th}}}{\sigma ^{-}}\otimes
q^{(1-n)\sigma ^{z}}\cdots \otimes \underset{m_{2}^{\text{th}}}{\sigma ^{-}}%
\otimes q^{(2-n)\sigma ^{z}}\cdots q^{-\sigma ^{z}}\otimes \underset{m_{n}^{%
\text{th}}}{\sigma ^{-}}\otimes 1\cdots \otimes 1
\end{multline*}
and 
\begin{multline*}
F_{1}^{(n)}=\lim_{\tilde{q}\rightarrow q}F_{1}^{n}(\tilde{q})/[n]_{\tilde{q}%
}!= \\
\tsum_{m_{i}}\,1\otimes \cdots 1\otimes \underset{m_{1}^{\text{th}}}{\sigma
^{-}}\otimes q^{-\sigma ^{z}}\cdots \otimes \underset{m_{2}^{\text{th}}}{%
\sigma ^{-}}\otimes q^{-2\sigma ^{z}}\cdots q^{-(n-1)\sigma ^{z}}\otimes 
\underset{m_{n}^{\text{th}}}{\sigma ^{-}}\otimes \cdots \otimes q^{-n\sigma
^{z}}
\end{multline*}
\begin{multline*}
F_{0}^{(n)}=\lim_{\tilde{q}\rightarrow q}F_{0}^{n}(\tilde{q})/[n]_{\tilde{q}%
}!= \\
\tsum_{m_{i}}\zeta _{m_{1}}^{-1}\cdots \zeta _{m_{n}}^{-1}\,1\otimes \cdots
1\otimes \underset{m_{1}^{\text{th}}}{\sigma ^{+}}\otimes q^{\sigma
^{z}}\cdots \otimes \underset{m_{2}^{\text{th}}}{\sigma ^{+}}\otimes
q^{2\sigma ^{z}}\cdots q^{(n-1)\sigma ^{z}}\otimes \underset{m_{n}^{\text{th}%
}}{\sigma ^{+}}\otimes \cdots \otimes q^{n\sigma ^{z}}
\end{multline*}
Here the sums are restricted to the $n$-tuples $(m_{1},...,m_{n})$ with $%
1\leq m_{1}<\cdots <m_{n}\leq M$. Remarkably, the subalgebra generated by
the above operators with powers equal to 
\begin{equation*}
n=N^{\prime }:=\left\{ 
\begin{array}{cc}
N, & N\;\text{odd} \\ 
N/2, & N\;\text{even}
\end{array}
\right.
\end{equation*}
is isomorphic to the ``classical'' loop algebra $U(\widetilde{sl}_{2})$ \cite
{CP,DFM,KM01}. The projection $U_{q}^{\text{res}}(\widetilde{sl}%
_{2})\rightarrow U(\widetilde{sl}_{2})$ is referred to as the quantum
Frobenius homomorphism \cite{CP}. In order to stress that this is an inf{i}%
nite-dimensional algebra we rewrite $U(\widetilde{sl}_{2})$ in terms of its
mode basis 
\begin{equation}
h_{m+n}=[x_{m}^{+},x_{n}^{-}],\;[h_{m},x_{n}^{\pm }]=\pm 2x_{m+n}^{\pm
},\;[h_{m},h_{n}]=0,\;[x_{m+1}^{\pm },x_{n}^{\pm }]=[x_{m}^{\pm
},x_{n+1}^{\pm }]\;.  \label{sl2}
\end{equation}
The generators $\{x_{m}^{\pm },h_{m}\}_{m\in \mathbb{Z}}$ can be
successively obtained from the Chevalley-Serre basis via the correspondence 
\cite{CP} 
\begin{equation}
E_{1}^{(N^{\prime })}\rightarrow x_{0}^{+},\quad F_{1}^{(N^{\prime
})}\rightarrow x_{0}^{-},\quad E_{0}^{(N^{\prime })}\rightarrow
x_{1}^{-},\quad F_{0}^{(N^{\prime })}\rightarrow x_{-1}^{+},\quad
2S^{z}/N^{\prime }\rightarrow h_{0}\;.  \label{gen}
\end{equation}
For later purposes let us identify the upper and lower Borel subalgebras $%
U(b_{\pm })\subset U(\widetilde{sl}_{2})$. In terms of the Chevalley-Serre
basis they are generated by $\{E_{0}^{(N^{\prime })},E_{1}^{(N^{\prime
})},2S^{z}/N^{\prime }\}$ and $\{F_{0}^{(N^{\prime })},F_{1}^{(N^{\prime
})},2S^{z}/N^{\prime }\}$, respectively. In the mode basis they simply
correspond to the algebras associated with the positive and negative
integers, 
\begin{equation}
U(b_{+})=\{x_{m}^{\pm },h_{m}\}_{m\in \mathbb{Z}_{>0}}\cup
\{x_{0}^{+},h_{0}\}\quad \text{and\quad }U(b_{-})=\{x_{m}^{\pm
},h_{m}\}_{m\in \mathbb{Z}_{<0}}\cup \{x_{0}^{-},h_{0}\}\;.  \label{Ubpm}
\end{equation}

We are now in the position to discuss the various symmetries of the twisted
six-vertex model at roots of unity. Taking the root-of-unity limit in (\ref
{Tb+}) and (\ref{Tb-}) we obtain the relations 
\begin{eqnarray}
E_{1}^{(N^{\prime })}T^{\lambda } &=&q^{N^{\prime }}T^{\lambda
}E_{1}^{(N^{\prime })}+\lambda ^{-\frac{1}{2}}(1-\lambda
K_{1})CE_{1}^{(N^{\prime }-1)}  \notag \\
E_{0}^{(N^{\prime })}T^{\lambda } &=&q^{N^{\prime }}T^{\lambda
}E_{0}^{(N^{\prime })}+\lambda ^{\frac{1}{2}}\,z(1-\lambda
^{-1}K_{0})BE_{0}^{(N^{\prime }-1)}  \label{Tb++}
\end{eqnarray}
and 
\begin{eqnarray}
F_{1}^{(N^{\prime })}T^{\lambda } &=&F_{1}^{(N^{\prime })}q^{N^{\prime
}}T^{\lambda }-\lambda ^{-\frac{1}{2}}q^{-N^{\prime }}(1-\lambda
K_{1}^{-1})F_{1}^{(N^{\prime }-1)}B  \notag \\
F_{0}^{(N^{\prime })}T^{\lambda } &=&F_{0}^{(N^{\prime })}q^{N^{\prime
}}T^{\lambda }-\lambda ^{\frac{1}{2}}\,z^{-1}q^{-N^{\prime }}(1-\lambda
^{-1}K_{0}^{-1})F_{0}^{(N^{\prime }-1)}C\;.  \label{Tb--}
\end{eqnarray}
Thus, upon inserting $K_{1}=K_{0}^{-1}=q^{2S^{z}}$ we now infer immediately
that whenever the terms in the brackets vanish we obtain a symmetry algebra.
For periodic boundary conditions, $\lambda =1$, we recover the previously
obtained loop algebra symmetry $U(\widetilde{sl}_{2})$ in the commensurate
sectors $2S^{z}=0\func{mod}N$ \cite{DFM}. For twisted boundary conditions
with $\lambda =q^{\mp n},\;0<n<N,$ we apparently only obtain ``half'' the
symmetry algebra, namely $U(b_{\pm })$, in the spin sectors $2S^{z}=\pm n%
\func{mod}N$. This is due to the fact that the Cartan generators $K_{i}$
appear with inverse powers in (\ref{Tb--}) compared to the ones in (\ref
{Tb++}). Obviously, we again recover the full loop algebra as a symmetry for
even roots of unity and $n=N^{\prime }$, i.e. the case of antiperiodic
boundary conditions $\lambda =q^{N^{\prime }}=-1$ discussed in \cite{D03}.

So far all discussed symmetries have only been established for certain
commensurate spin-sectors. If we choose, however, the twist parameter to
depend on the total spin the inf{i}nite-dimensional non-abelian algebras (%
\ref{Ubpm}) extend to a symmetry for \emph{all} spin sectors. Namely, we now
consider the transfer matrices 
\begin{equation}
T^{\pm }(z)=\limfunc{Tr}_{0}q^{\pm \sigma ^{z}\otimes S^{z}}R_{0M}(z/\zeta
_{M})\cdots R_{01}(z/\zeta _{1})=q^{\pm S^{z}}A(z)+q^{\mp S^{z}}D(z)\;.
\label{Tpm}
\end{equation}
At f{i}rst sight one might be worried that the twist parameter is now an
operator instead of a mere constant. But according to (\ref{KABCD}) we have $%
[A,q^{S^{z}}]=[D,q^{S^{z}}]=0$ whence upon employing the standard relations
of the six-vertex Yang-Baxter algebra the integrability of the model is
ensured, i.e. 
\begin{equation}
\lbrack T^{\pm }(z),T^{\pm }(w)]=[A(z),D(w)]+[D(z),A(w)]=0\;.
\end{equation}
Thus, all results generalise in a straightforward manner to this case. The
only difference is that in the commutation relations (\ref{Tb+}) and (\ref
{Tb-}) we now collect additional factors $\tilde{q}^{\pm n}$ on the left
hand side of the equations as we have to ``pull'' $\tilde{q}^{\pm S^{z}}$
past the generators $E_{i}^{n},F_{i}^{n}$, 
\begin{eqnarray}
E_{1}^{n}(\tilde{q}^{-n-S^{z}}A+\tilde{q}^{n+S^{z}}D) &=&(\tilde{q}%
^{n-S^{z}}A+\tilde{q}^{-n+S^{z}}D)E_{1}^{n}  \notag \\
F_{1}^{n}(\tilde{q}^{-n+S^{z}}A+\tilde{q}^{n-S^{z}}D) &=&(\tilde{q}%
^{n+S^{z}}A+\tilde{q}^{-n-S^{z}}D)F_{1}^{n}\;.
\end{eqnarray}
The relations for the aff{i}ne step operators follow from (\ref{auto}). As a
consequence the transfer matrices $T^{\pm }$ now always commute with the
generators of (\ref{Ubpm}) in the root of unity limit $\tilde{q}\rightarrow
q $ (instead of anticommuting for even roots of unity cf. equations (\ref
{Tb++}) and (\ref{Tb--})), 
\begin{equation}
\lbrack T^{+}(z),U(b_{-})]=0\quad \text{and\quad }[T^{-}(z),U(b_{+})]=0\;.
\end{equation}
These symmetries hold for all spin sectors and are the main result of this
letter. Note that the case of periodic boundary conditions \cite{DFM} is
contained in these models for the sectors $2S^{z}=0\func{mod}N$, where both
transfer matrices coincide and the symmetry is enhanced to the full loop
algebra.

\section{Conclusions}

Let us summarize the established symmetry algebras for the twisted
inhomogeneous six-vertex model and their corresponding commensurate sectors
in the following table,\medskip

\begin{center}
\begin{tabular}{|l|l|l|}
\hline\hline
twist & symmetry & spin-sector \\ \hline\hline
$\lambda =1$ & $U(\widetilde{sl}_{2})$ & $2S^{z}=0\func{mod}N$ \\ 
\hline\hline
$\lambda =q^{n}$ & $U(b_{\pm })$ & $2S^{z}=\mp n\func{mod}N$ \\ \hline\hline
$\lambda =q^{\mp 2S^{z}}$ & $U(b_{\pm })$ & all sectors \\ \hline\hline
\end{tabular}
\medskip

{\small Table 1. The various symmetry algebras for the twisted six-vertex
model at a primitive root of unity }$q^{N}=1${\small \ and the spin-sectors
in which they have been constructed explicitly.}\medskip
\end{center}

\noindent Note that the above f{i}ndings do not exclude the possibility that
the symmetries found \cite{DFM,D03} for the boundary conditions $\lambda
\neq q^{\pm 2S^{z}}$ can be extended to all spin-sectors as well. For
periodic boundary conditions $\lambda =1$ it has been argued in \cite{DFM}
that one might have to use projection operators to obtain the symmetry
algebra in the incommensurate sectors. As mentioned in the introduction this
has been explicitly demonstrated at the free fermion point, i.e. the XX
model ($N/2=N^{\prime }=2$). For $N^{\prime }\geq 2$ it has been proven that
operators of the type $E_{1}^{n}E_{0}^{n},E_{1}^{N^{\prime
}-n}E_{0}^{N^{\prime }-n}$ etc. commute with the transfer matrix when $%
\lambda =1$ and $2S^{z}=2n\func{mod}N$, cf. equation (3.42) and Section 3.5,
Appendix A.5 in \cite{DFM}. In equation (3.43) of the same work eight
operators are stated which should (anti)commute with the transfer matrix in
the incommensurate sector $2S^{z}=2n\func{mod}N$ and a numerical procedure
is described how the loop algebra relations have been verified for $%
N^{\prime }=3$.\smallskip

For the transfer matrices (\ref{Tpm}) we obviously do not need any
projection operators to extend the symmetry to all spin-sectors which
indicates that these models possess a higher level of degeneracies in their
spectrum compared to the other boundary conditions. That this is indeed the
case has been numerically verif{i}ed in the spin-sectors $S^{z}=2,-1$ of the 
$M=6$ spin-chain when $q^{3}=1$; see Graph 1 and Graph 2. Furthermore, our
results for the twisted case when $\lambda \neq q^{\pm 2S^{z}}$ suggest that
in the incommensurate sectors one might also encounter a smaller symmetry
algebra as the spin-sector $2S^{z}=0\func{mod}N$ is clearly
distinguished.\smallskip

We emphasize again that in comparison with previously established
non-abelian symmetries, e.g. the \emph{f{i}nite} quantum group symmetry $%
U_{q}(sl_{2})$ for the chain with open boundary conditions \cite{PS}, the
symmetries established here involve \emph{inf{i}nite}-dimensional algebras
which impose more powerful restrictions. The next step in this context is to
relate the representation theory of these algebras to the Bethe
ansatz.\smallskip

For periodic boundary conditions $\lambda =1$ this has already partially
been done in \cite{FM01b,D02,KQ,KQ2}. In \cite{FM01b} creation operators
involving complete strings have been constructed which involve two
polynomials depending on the Bethe roots. Based on numerical results one of
these polynomials has been conjectured \cite{FM01a,FM01b} to coincide with
the classical limit ($q\rightarrow 1$) of the Drinfeld polynomial \cite{Drin}
which describes the irreducible representations of the loop algebra \cite{CP}
in the sectors $2S^{z}=0\func{mod}N$. The previously formulated conjecture 
\cite{DFM,FM01a,FM01b} that the regular XXZ Bethe vectors correspond to the
highest weight vectors of the loop algebra has been investigated in \cite
{D02} by means of the algebraic Bethe ansatz. Also here the results have
been limited to the commensurate sectors $2S^{z}=0\func{mod}N$ where the
algebraic structure of the symmetry generators has been identif{i}ed. In 
\cite{KQ,KQ2} the degeneracies of the periodic six-vertex model have been
investigated from a different point of view by applying representation
theory to construct analogues of Baxter's $Q$-operator. In \cite{KQ2} the
classical Drinfeld polynomial has been identif{i}ed in the spectrum of these 
$Q$-operators for several explicit examples when $N=3$.\smallskip

Clearly, the advantage of imposing the quasi-periodic boundary conditions $%
\lambda =q^{\pm 2S^{z}}$ is that the symmetry algebra is now known for all
spin-sectors while at the same time leaving the algebraic structure of the
Bethe ansatz largely unchanged. This makes the twisted model (\ref{Tpm}) an
ideal candidate for representation theoretic investigations and one can
expect to f{i}nd similar results as for the periodic case. Of particular
interest in this context is also the study of f{i}nite-size effects in the
thermodynamic limit, similar to those done in existing numerical
investigations of the twisted XXZ chain e.g. \cite{ABB88,SS90,YF92,RES95}.
These issues will be addressed in a forthcoming paper \cite{future}%
.\smallskip

F{i}nally, it needs to be pointed out that the proof of the inf{i}%
nite-dimensional symmetries given in this article has only made use of the
intertwining property of the monodromy matrix. This property is common to a
large class of integrable vertex models associated with trigonometric
solutions to the Yang-Baxter equation and quantum aff{i}ne (super)algebras.
Despite the obvious modif{i}cations in the algebraic structure of the
Yang-Baxter algebra we expect that the results found here can be extended to
these models similar as the periodic case has been generalised to other
models in \cite{KM01} and \cite{KR02} (albeit with different methods).%
{\small \medskip }

\noindent \textbf{Acknowledgments}. Part of this work has been done during a
research stay at Bergische Universit\"{a}t Wuppertal and the author would
like to thank the group of Andreas Kl\"{u}mper for their hospitality, in
particular Frank G\"{o}hmann for his kind invitation. The author is also
grateful to Harry Braden, Tetsuo Deguchi and Barry McCoy for comments on a
draft version of this letter. This work has been f{i}nancially supported by
the EPSRC Grant GR/R93773/01.{\small \medskip }

\begin{center}
\includegraphics[totalheight=11cm]{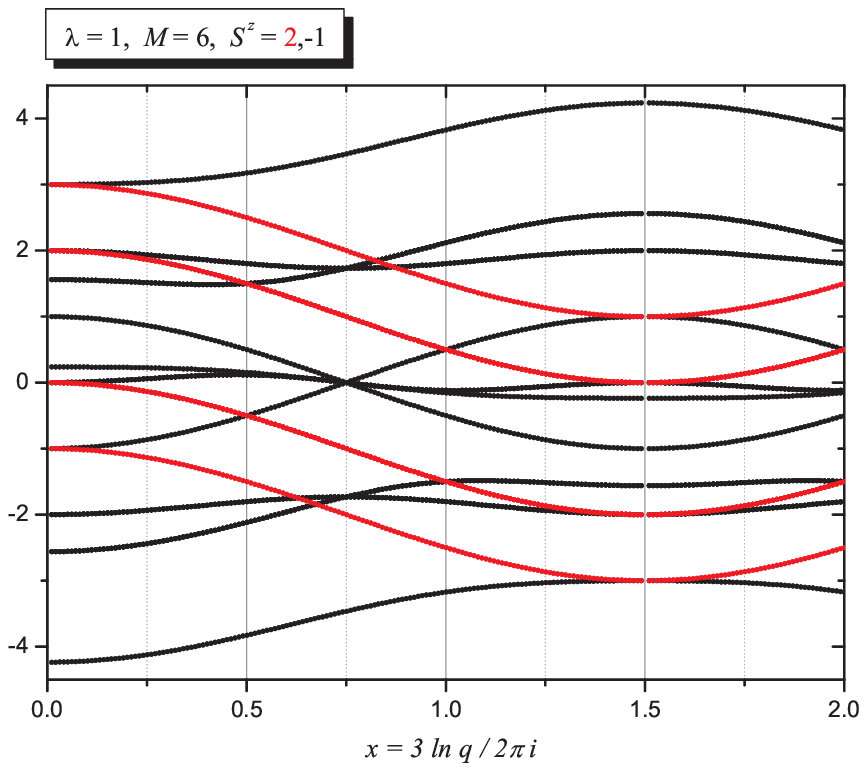}
\end{center}

{\small Graph 1. The eigenvalues for the periodic XXZ Hamiltonian (\ref{H})
with $M=6$ as a function of the deformation parameter }$q=\exp 2\pi ix/3$%
{\small . The eigenvalues for the spin sector }$S^{z}=2${\small \ are shown
in red colour. The eigenvalues corresponding to the two inner lines are each
doubly degenerate. The eigenvalues corresponding to the spin }$S^{z}=-1$%
{\small \ sector are shown in black colour, also here some of them are
doubly degenerate. At the root-of-unity points }$x=1/2,1${\small \ we see
that additional degeneracies occur between eigenvalues from the two
different spin sectors. Note that these are incommensurate sectors. The
distinguished points }$x=3/4,3/2${\small \ correspond to the XX model and
the case when }$q=-1${\small .}\bigskip

\begin{center}
\includegraphics[totalheight=11cm]{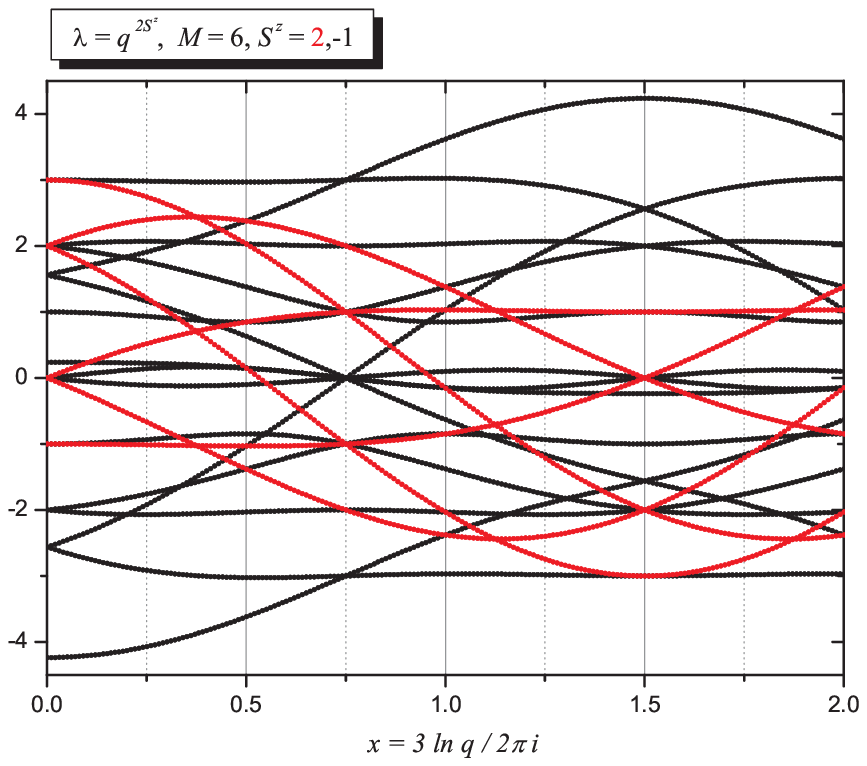}
\end{center}

{\small Graph 2. The eigenvalues for the twisted XXZ Hamiltonian (\ref{Ham})
with the twist depending on the spin $\lambda=q^{2S^z}$. As in the periodic
case the eigenvalues in the spin-sector }$S^{z}=2${\small \ and }$S^{z}=-1$%
{\small \ are displayed in red and black, respectively. Unlike in the
periodic case the degeneracies of the Hamiltonian within the respective spin
sectors are lifted. In addition, we see that at the root-of-unity values }$%
x=1/2,1${\small \ now all six eigenvalues of the }$S^{z}=2${\small \ sector
become degenerate with eigenvalues in the }$S^{z}=-1$ {\small sector. These
degeneracies indicate the discussed }$U(b_{\pm })$ {\small symmetries.}%

\end{document}